\documentclass[12pt]{iopart}
\usepackage{epsfig}
\usepackage{bm}
\usepackage[dvipdfmx]{hyperref}

\newcommand{\be}{\begin{eqnarray}}
\newcommand{\ee}{\end{eqnarray}}
\newcommand{\bmat}{\left(\begin{array}}
\newcommand{\emat}{\end{array}\right)}
\newcommand{\no}{\nonumber}

\begin{document}

\title{Adiabatic theorem for classical stochastic processes}

\author{Kazutaka Takahashi}

\address{Department of Physics Engineering, Faculty of Engineering,
Mie University, Mie 514--8507, Japan}
\ead{ktaka@phen.mie-u.ac.jp}
\vspace{10pt}
\begin{indented}
\item[]March 2024
\end{indented}

\begin{abstract}
We apply adiabatic theorems developed for quantum mechanics
to stochastic annealing processes 
described by the classical master equation with
a time-dependent generator.
When the instantaneous stationary state is unique
and the minimum decay rate $g$ is nonzero, 
the time-evolved state is basically relaxed to 
the instantaneous stationary state.
By formulating an asymptotic expansion rigorously,
we derive conditions for the annealing time $\tau$
that the state is close to the instantaneous stationary state.
Depending on the time dependence of the generator,
typical conditions are written as 
$\tau> \mathrm{const.}\times g^{-\alpha}$
with $1\le \alpha \le 2$.
We also find that a rigorous treatment gives the scaling
$\tau>\mathrm{const.}\times g^{-2}|\ln g|$.
\end{abstract}

%
%
%
%
%

\section{Introduction}

One of efficient methods for changing a state to a desired state 
is to control the system slowly by an external operation.
The system itself evolves by a time evolution law and 
the problem is represented by a differential equation
for a nonautonomous system.
Since the stationary state changes as a function of time,
it is generally hard to find the solution of the differential equation.
When the system is changed very slowly,
we rely on the adiabatic approximation.
The adiabatic theorem of quantum mechanics describes
how the evolved state deviates from the ideal one~\cite{Born28, Kato50}.
Rigorous treatments of the theorem 
have a long history and we can find many variations depending on
settings~\cite{Avron87, Nenciu89, Nenciu93, Avron99, Jansen07, Lidar09, Elgart12, Venuti16}.
Those studies revealed that the naive condition discussed
in a standard textbook on quantum mechanics is not necessarily correct.
A careful analysis of the asymptotic expansion gives
nontrivial contributions.

One of the most prominent applications of the adiabatic theorem is
quantum annealing/adiabatic quantum
computation~\cite{Apolloni89, Somorjai91, Amara93, Finnila94, Farhi98, Kadowaki98, Brooke99, Farhi00, Farhi01,  Albash18}.
When we want to know the ground state of an Ising-spin Hamiltonian,
we drive the system by a transverse field.
The adiabatic theorem basically evaluates 
the annealing time that suppresses the error.
The minimum annealing time is determined from
the change rate of the Hamiltonian and  
the instantaneous energy gap between
the ground state and the excited states.

Historically, the idea of the quantum annealing stemmed from
a classical optimization method called simulated annealing~\cite{Kirkpatrick83}.
The stationary state is described by the Gibbs distribution and 
we decrease the temperature of the system slowly to
find the optimized configuration of the free energy.
In that case, Geman--Geman found a protocol that guarantees
a convergence to the optimized solution~\cite{Geman84}.
As a function of a dimensionless time $t$
ranging from 0 to $\infty$,
the temperature $T$ is decreased as $T(t)\propto 1/\log(t+1)$.

The simulated annealing is a classical process.
When the dynamics is assumed to be Markovian, the time evolution
is generally described by the classical master equation.
Since the master equation is formally equivalent to 
the imaginary-time Schr\"odinger equation,
it is not difficult to apply the adiabatic theorems
developed for the case of the Schr\"odinger equation~\cite{Morita08}.
However, the imaginary time gives a relaxation dynamics,
which implies that the process is not strongly dependent
on the history of the time evolution.
Rigorous treatments of the adiabatic theorem in quantum mechanics
give contributions which are represented by integrals over the whole time.
We expect that such contributions are absent in
the classical stochastic processes
and the theorem is greatly simplified.
In the classical master equation with a time-dependent generator, 
we need to discuss how the relaxation and annealing dynamics 
affect the state of the system.

The classical master equation treats a time evolution of
a probability distribution.
Although the equation is formally equivalent to the imaginary-time 
Schr\"odinger equation, the probability nature restricts
possible patterns of dynamics.
As a result, we expect that we can find some simplifications on 
the adiabatic theorem.

In the present study, we treat the case
where the instantaneous stationary state is defined uniquely.
This corresponds to the standard setting
used in the quantum cases~\cite{Kato50}.
We do not assume explicit form of the stationary state such
as the Gibbs distribution 
and aim at finding a general adiabatic theorem
which is derived under several fundamental conditions described below.

The classical master equation can be described by a reduction
of the quantum master equation.
A general formulation of completely-positive trace-preserving
map~\cite{Nielsen00}
is proved to be useful for describing quantum and classical dynamics
in a unified way.
The adiabatic theorem for the quantum master equation is formally described
by the Liouvillian superoperator~\cite{Venuti16}.
However, the physical picture of the result is not so obvious and
it is hard to obtain the corresponding result at the classical limit.
We note that 
the decomposition of the quantum master equation
into quantum and classical parts is a nontrivial task~\cite{Funo19}.
Also, we are interested in purely classical processes
such as the simulated annealing.

The organization of the paper is as follows.
In Sec.~\ref{sec:intro}, we formulate the problem
and describe settings used throughout the present study.
We introduce adiabatic dynamics in Sec.~\ref{sec:ad}.
It is used to develop adiabatic theorems in Sec.~\ref{sec:th}.
In Sec.~\ref{sec:ex}, 
we treat several examples to examine general results.
The last section~\ref{sec:conc} is devoted to conclusion.

\section{System}
\label{sec:intro}

In stochastic dynamical processes, the state of the system is specified by
a time-dependent probability distribution.
We assume that the dynamics is a Markov process and
the transition rate is inhomogeneous.
We introduce a scaled time $s=t/\tau$ where $t$ is the physical time and
$\tau$ is the annealing time.
The time evolution is carried out from $t=0$ to $t=\tau$.
Correspondingly, $s$ runs from 0 to 1.
The probability distribution denoted by 
$|p_\tau(s)\rangle=\sum_{n=0}^{N-1}|n\rangle\langle n|p_\tau(s)\rangle$, 
where $\langle n|p_\tau(s)\rangle$ represents the $n$th component of
the probability distribution and $N$ is the number of events, 
obeys the master equation 
\be
 |\dot{p}_\tau(s)\rangle = \tau W(s)|p_\tau(s)\rangle.
\ee
$W(s)$ represents the transition-rate matrix.
We denote the derivative with respect to the scaled time $s$
by the dot symbol.
Since the probability nature must be maintained throughout the time evolution,
each offdiagonal component of the transition-rate matrix
is nonnegative and the matrix satisfies
\be
 \langle L_0|W(s)=0,
\ee
where $\langle L_0|=\sum_n\langle n|$ with $\langle m|n\rangle=\delta_{m,n}$.

Throughout this study, we assume that there exists a unique 
instantaneous stationary state $|p^{(\mathrm{st})}(s)\rangle$
defined from the relation 
\be
 W(s)|p^{(\mathrm{st})}(s)\rangle =0.
\ee
When we operate the system slowly,
the state of the system is relaxed to the instantaneous stationary state.
For simplicity, we set that the initial probability distribution
$|p_\tau(0)\rangle=|p_0\rangle$ 
is given by the stationary state 
with respect to the initial transition-rate matrix $W(0)=W_0$.
We are interested in a long-time behavior and
the final state is expected to be insensitive
to the choice of the initial state.

The decay rate is characterized by the instantaneous eigenvalues of $W(s)$.
When $N$ is finite, we can use the spectral representation
\be
 W(s)=\sum_{n=0}^{N-1}\Lambda_n(s)|R_n(s)\rangle\langle L_n(s)|, \label{sp}
\ee
where the left and right eigenstates satisfy
$\langle L_m(s)|R_n(s)\rangle=\delta_{m,n}$
and $\sum_n |R_n(s)\rangle\langle L_n(s)|=1$.
The stationary state is denoted by the component $n=0$.
That is, we have 
$|p^{(\mathrm{st})}(s)\rangle = |R_0(s)\rangle$,
$|p_0\rangle=|p^{(\mathrm{st})}(0)\rangle=|R_0(0)\rangle$,
and $\Lambda_0(s)=0$.
The assumption of the unique stationary state denotes
that the minimum decay rate is positive:
\be
 g(s)=\min_{n\ne 0}|\mathrm{Re}\,\Lambda_n(s)|>0. \label{gap}
\ee

The time-evolved state $|p_\tau(s)\rangle$ is different from
the instantaneous stationary state $|p^{(\mathrm{st})}(s)\rangle$.
One of the standard quantities for measuring the deviation 
is given by the trace distance 
\be
 d_\tau(s)=\frac{1}{2}\sum_{n}|
 \langle n|p_\tau(s)\rangle-\langle n|p^{(\mathrm{st})}(s)\rangle|.
 \label{tracedist}
\ee
When the annealing time $\tau$ is large enough,
$|p_\tau(s)\rangle$ is close to $|p^{(\mathrm{st})}(s)\rangle$.
The main problem discussed in the present study is
to estimate the magnitude of $\tau$ that results in a small $d_\tau(s)$.

The master equation basically describes relaxation dynamics.
When the time-evolved state is different from a fixed stationary state,
it is relaxed to the stationary state.
The decay rate can be estimated by $g(s)$.
The state is relaxed to the stationary state immediately
if the annealing time satisfies the condition
\be
 \tau \ge \mathrm{const.}\times \frac{1}{g(s)}.
 \label{g1}
\ee
In the present case, the stationary state 
is slowly varied as a function of time and 
we need to discuss adiabatic dynamics.
If we naively apply the simplest version of the quantum adiabatic theorem
to the present system, 
the condition for a small deviation from the adiabatic state
is given by 
\be
 \tau \ge \mathrm{const.}\times\frac{\|\dot{W}(s)\|}{g^2(s)},
 \label{g2}
\ee
where $\|\cdot\|$ is a proper norm of matrices.
Since we use the transition-rate matrix for the generator of
the time evolution instead of the Hamiltonian,
the minimum decay rate $g(s)$ plays the role of the energy gap.
To develop the adiabatic theorem for this system,
we discuss the scaling of the annealing time at $g(s)\to 0$.
In that case, 
the relaxation condition in Eq.~(\ref{g1}) is automatically satisfied 
when the naive adiabatic condition in Eq.~(\ref{g2}) holds, and
we can focus on the adiabatic condition in a small-$g$ regime.

A careful analysis of the adiabatic theorem in quantum mechanics
indicates that the condition corresponding to Eq.~(\ref{g2})
is not necessarily correct~\cite{Jansen07, Elgart12}.
The main aim of the present study is to find a condition 
that is valid in the present system.
We examine an asymptotic behavior of the trace distance.
It is expected to be written as  
\be
 d_\tau(s)\sim\frac{\tau_1(s)}{\tau}+\frac{\tau_2(s)}{\tau^2}+\cdots
 +e^{-\tau \int_0^s ds'\,g(s')}(\cdots). \label{asymp}
\ee
In the regime we are mainly interested in, 
the last exponential term is negligibly small 
and the adiabatic condition is obtained from the power-law part.
Each term of the expansion is bounded from above and we estimate 
a minimum annealing time such that
the distance is bounded by a specified maximum error
as $d_\tau(s)\le\delta$.

\section{Adiabatic dynamics}
\label{sec:ad}

The stationary state $|p^{(\mathrm{st})}(s)\rangle$ has nothing to do
with real-time dynamics and we introduce a virtual dynamical process
that results in $|p^{(\mathrm{st})}(s)\rangle$.
Differentiating the stationary state with $s$, we can write 
\be
 |\dot{p}^{(\mathrm{st})}(s)\rangle
 =\left(\tau W(s)+\dot{P}(s)\right)|p^{(\mathrm{st})}(s)\rangle,
 \label{pst}
\ee
where $P(s)$ represents the projection onto the stationary state:
\be
 P(s)=|R_0(s)\rangle\langle L_0|. \label{projection}
\ee
Since the left eigenstate $\langle L_0|$ is time independent,
we can write 
\be
 \dot{P}(s)=\dot{P}(s)P(s)=Q(s)\dot{P}(s)P(s), \label{pdot}
\ee
where $Q(s)=1-P(s)$.
This simple structure of the projection operator is
one of the major differences from the quantum case~\cite{Kato50}.
We have introduced $W(s)$ in Eq.~(\ref{pst}) for later convenience.
Since $W(s)|p^{(\mathrm{st})}(s)\rangle=0$,
it does not give any contribution to the stationary state.

Generalizing this time-evolution law to arbitrary states, 
we introduce the time-evolution operator $U_\tau(s,s')$.
It obeys the equation of motion 
\be
 \partial_s U_\tau(s,s')=
 \left(\tau W(s)+\dot{P}(s)\right)U_\tau(s,s'), 
\ee
with the boundary condition $U_\tau(s,s)=1$
and the associative law $U_\tau(s,s')U_\tau(s',s'')=U_\tau(s,s'')$.
We can write
$|p^{(\mathrm{st})}(s)\rangle=U_\tau(s,0)|p_0\rangle$ and 
\be
 U_\tau(s,s')P(s')=P(s)U_\tau(s,s'). \label{uppu}
\ee
This relation can be verified by applying the time derivative to
this expression.

We use a Volterra integral form to derive the adiabatic theorem.
Generally, we can write for two kinds of time-evolved states
$|p^{(1)}(t)\rangle$ and $|p^{(2)}(t)\rangle$ as 
\be
 |p^{(1)}(t)\rangle-|p^{(2)}(t)\rangle
 =\int_0^t dt'\,U^{(2)}(t,t')(W^{(1)}(t')-W^{(2)}(t'))
 |p^{(1)}(t')\rangle, 
\ee
where $W^{(1)}(t)$ is the generator for the first state $|p^{(1)}(t)\rangle$,
$W^{(2)}(t)$ is for the second, and
$U^{(1)}(t,t')$ is the time-evolution operator of the first state.
Applying this general formula to the present case, we have
\be
 |p_\tau(s)\rangle-|p^{(\mathrm{st})}(s)\rangle
 &=& -\int_0^s ds'\,U_\tau(s,s')\dot{P}(s')
 |p_\tau(s')\rangle \no\\
 &=& -\int_0^s ds'\,U_\tau(s,s')|\dot{p}^{(\mathrm{st})}(s')\rangle.
 \label{zero}
\ee
Remarkably, the last expression 
is independent of $|p_\tau(s)\rangle$.
This is due to the property in Eq.~(\ref{pdot}).

We note that $U_\tau(s,s')$ is the time evolution operator
of the stationary state $|p^{(\mathrm{st})}(s)\rangle$ and
not of the time-evolved state $|p_\tau(s)\rangle$.
In quantum mechanics, the corresponding operator was introduced
in Ref.~\cite{Kato50}.
Also, $U_\tau(s,s')$ is different from the time evolution operator 
of the adiabatic state defined from
\be
 \partial_s U_\tau^{(\mathrm{ad})}(s,s')=
 \left(\tau W(s)+W^{(\mathrm{cd})}(s)\right)U_\tau^{(\mathrm{ad})}(s,s'), 
\ee
where 
\be
 W^{(\mathrm{cd})}(s)=\sum_n (1-|R_n(s)\rangle\langle L_n(s)|)
 |\dot{R}_n(s)\rangle\langle L_n(s)|.
\ee
When we use this time-evolution law,
arbitrary initial states written as 
\be
 |p_0\rangle = \sum_n c_n |R_n(0)\rangle,
\ee
are transformed to the adiabatic state 
\be
 U_\tau^{(\mathrm{ad})}(s,0)|p_0\rangle
 &=& \sum_n c_n |R_n(s)\rangle
 \no\\ &&\times\exp\left(\tau\int_0^s ds'\,\Lambda_n(s')
 -\int_0^s ds'\,\langle L_n(s')|\dot{R}_n(s')\rangle\right).
\ee
$W^{(\mathrm{cd})}(s)$ is the counterpart of the counterdiabatic term defined
in quantum mechanics~\cite{Demirplak03, Demirplak05, Berry09}
and was used in classical stochastic
processes~\cite{Takahashi20, Funo20, Takahashi23}.
The generator $\dot{P}(s)$ for $U_\tau(s,s')$ is obtained from
$W^{(\mathrm{cd})}(s)$ as $\dot{P}(s)=Q(s)W^{(\mathrm{cd})}(s)P(s)$.
The driving by $U_\tau^{(\mathrm{ad})}$
with the initial condition $|p_0\rangle=|R_0(0)\rangle$
gives the identical time evolution as that by $U_\tau$.
Since we can use some useful properties described by the projection operator 
such as the last expression in Eq.~(\ref{zero}),
we use $U_\tau$ rather than $U_\tau^{(\mathrm{ad})}$.

\section{Adiabatic theorem}
\label{sec:th}

\subsection{Asymptotic expansion}

The integral in Eq.~(\ref{zero}) is written as 
\be
 -\int_0^s ds'\,U_\tau(s,s')|\dot{p}^{(\mathrm{st})}(s')\rangle
 =-\int_0^s ds'\,Q(s)U_\tau(s,s')Q(s')|\dot{p}^{(\mathrm{st})}(s')\rangle.
\ee
That is, the time-evolution operator $U_\tau$ in Eq.~(\ref{zero}) 
acts on the projected space excluding the stationary state.
To find an asymptotic form of this expression,
we introduce
\be
 |\phi^{(1)}(s)\rangle = G(s)\partial_s|p^{(\mathrm{st})}(s)\rangle,
\ee
where 
\be
 G(s) = Q(s)\frac{1}{-W(s)}Q(s).
\ee
Equation (\ref{zero}) is written as 
\be
 |p_\tau(s)\rangle-|p^{(\mathrm{st})}(s)\rangle
 =\int_0^s ds'\,U_\tau(s,s')W(s')|\phi^{(1)}(s')\rangle.
\ee
By noting 
\be
 \partial_{s'}\left(U_\tau(s,s')|\phi^{(1)}(s')\rangle\right)
 =-\tau U_\tau(s,s')W(s')|\phi^{(1)}(s')\rangle
 +U_\tau(s,s')|\dot{\phi}^{(1)}(s')\rangle, \no\\ 
\ee
we rewrite the integral as 
\be
 |p_\tau(s)\rangle-|p^{(\mathrm{st})}(s)\rangle
 &=& -\frac{1}{\tau}\left(
 |\phi^{(1)}(s)\rangle-U_\tau(s,0)|\phi^{(1)}(0)\rangle\right)
 \no\\ &&
 +\frac{1}{\tau}\int_0^s ds'\,
 U_\tau(s,s')|\dot{\phi}^{(1)}(s')\rangle. \label{first}
\ee
The last term has a similar form as
the integral in Eq.~(\ref{zero}) and we can apply similar transformations
repeatedly.
We introduce 
\be
 |\phi^{(k)}(s)\rangle = G(s)\partial_s|\phi^{(k-1)}(s)\rangle
 = \left(G(s)\partial_s\right)^k|p^{(\mathrm{st})}(s)\rangle,
\ee
for an integer $k$ to write 
\be
 |p_\tau(s)\rangle-|p^{(\mathrm{st})}(s)\rangle
 &=& \sum_{k=1}^M\left(\frac{-1}{\tau}\right)^k\left(
 |\phi^{(k)}(s)\rangle-U_\tau(s,0)|\phi^{(k)}(0)\rangle\right)
 \no\\ 
 && -\left(\frac{-1}{\tau}\right)^M\int_0^s ds'\,
 U_\tau(s,s')|\dot{\phi}^{(M)}(s')\rangle, \label{Mth}
\ee
where $M$ is an arbitrary integer.

This technique is essentially the same as that used in quantum
adiabatic theorems~\cite{Avron87, Avron99, Jansen07}.
The projection operator is written in an integral form as 
\be
 P(s)=\oint \frac{dz}{2\pi i}\,\frac{1}{z-W(s)}.
\ee
The integral contour in the complex-$z$ plane 
encloses the origin and the other eigenvalues
of $W(s)$ denoted by points in the complex plane are outside the contour.
We note that the condition $g(s)>0$ is required
to make the integral well-defined.
In a similar way, we can define a transformation 
\be
 \tilde{X}(s)=\oint \frac{dz}{2\pi i}\,
 \frac{1}{z-W(s)}X(s)\frac{1}{z-W(s)},
\ee
for an arbitrary operator $X(s)$.
We can show that
$\Phi^{(k)}(s)=\tilde{\dot{\Phi}}^{(k-1)}(s)$ with $\Phi^{(0)}(s)=P(s)$
is written as 
$\Phi^{(k)}(s)=|\phi^{(k)}(s)\rangle\langle L_0|$ 
with $|\phi^{(0)}(s)\rangle=|R_0(s)\rangle=|p^{(\mathrm{st})}(s)\rangle$.
Due to the property in Eq.~(\ref{pdot}), 
the resulting form takes a considerably simpler form
compared to the quantum case.

Equation (\ref{Mth}) consists of three parts.
The first part is a simple expansion  with respect to $1/\tau$
and each term is characterized by $|\phi^{(k)}(s)\rangle$.
The second part involves $U_\tau(s,0)$.
It acts on a projected space as $Q(s)U_\tau(s,0)Q(0)$,
which implies that it involves exponentially-decaying factors
like $\exp(-\tau \int_{0}^s ds'\,g(s'))$.
Then, $U_\tau(1,0)|\phi^{(k)}(0)\rangle$ is negligibly small
for $\tau \int_0^1 ds\,g(s)\gg 1$.
This property is reasonable
since the final state must be insensitive to the choice of the initial
state, if the ergodicity condition is satisfied~\cite{Morita08}.
The last part is an integral over the whole time period and
represents the correction
when the expansion in the first part is truncated at a finite $M$.

It is instructive to compare the asymptotic form in Eq.~(\ref{Mth})
to that found in the case of the quantum master equation~\cite{Venuti16}.
The major difference is that, at each order $k$, the contribution represented 
by in integral over the whole time period is absent.
Although the absence of the integral is physically reasonable
from the above discussion on the time-evolution operator,
it is not trivial to find that such terms never appear in the expression.

\subsection{Bounds of expansion coefficients (1)}

We can perform the expansion in Eq.~(\ref{Mth}) up to a desired order $M$.
Truncating the expansion is accomplished by neglecting the last term
in Eq.~(\ref{Mth}).
In rigorous treatments of the adiabatic theorems,
the last term contribution is kept to derive a nontrivial result.

In this subsection, by taking $M\to\infty$ formally, we examine 
\be
 |p_\tau(s)\rangle-|p^{(\mathrm{st})}(s)\rangle
 \sim \sum_{k=1}^\infty\left(\frac{-1}{\tau}\right)^k|\phi^{(k)}(s)\rangle.
\ee
When we use the trace distance in Eq.~(\ref{tracedist}) as a measure,
it is bounded by using the relation 
\be
 \frac{1}{2}\sum_{n=0}^{N-1} |\langle n|\phi\rangle|
 =\frac{1}{2}\sum_n 
 \mathrm{sgn}\,(\langle n|\phi\rangle)\cdot\langle n|\phi\rangle
 \le \frac{1}{2}\sqrt{N\sum_n(\langle n|\phi\rangle)^2}.
\ee
We obtain
\be
 d_\tau(1)\sim \frac{1}{2}\sum_{n=0}^{N-1}\left|
 \sum_{k=1}^\infty\left(\frac{-1}{\tau}\right)^k\langle n|\phi^{(k)}(1)\rangle
 \right|
 \le \frac{\sqrt{N}}{2}\sum_{k=1}^\infty \frac{1}{\tau^k}
 \||\phi^{(k)}(1)\rangle\|,
\ee
where $\|\cdot\|$ denotes the vector norm.
We use $\left\|c_1|\phi_1\rangle+c_2|\phi_2\rangle\right\|\le
|c_1|\left\||\phi_1\rangle\right\|+|c_2|\left\||\phi_2\rangle\right\|$.

To evaluate the norm of each term, we need to know
explicit forms of $|\phi^{(k)}(s)\rangle$.
Using the relation
$W(s)|\dot{p}^{(\mathrm{st})}(s)\rangle=-\dot{W}(s)|p^{(\mathrm{st})}(s)\rangle$,
we write $|\phi^{(1)}(s)\rangle$ as 
\be
 |\phi^{(1)}(s)\rangle = G^2(s)\dot{W}(s)|p^{(\mathrm{st})}(s)\rangle.
\ee
This form has a bound at $s=1$ as 
\be
 \||\phi^{(1)}(1)\rangle\|\le \frac{\|\dot{W}(1)\|}{g^2(1)},
\ee
where $\|\cdot\|$ on the right hand side denotes the operator norm.
Similarly, we have 
\be
 |\phi^{(2)}(s)\rangle
 &=& \left(2G^3(s)\dot{W}(s)G(s)\dot{W}(s)
 +G^2(s)\dot{W}(s)G^2(s)\dot{W}(s)
 \right. \no\\ && \left.
 +G^3(s)\ddot{W}(s)
 \right)|p^{(\mathrm{st})}(s)\rangle,
\ee
and 
\be
 \||\phi^{(2)}(1)\rangle\|\le 3\frac{\|\dot{W}(1)\|^2}{g^4(1)}
 +\frac{\|\ddot{W}(1)\|}{g^3(1)}.
\ee
These calculations indicate that
at each order $k$
the leading term is proportional to $(\|\dot{W}(s)\|/g^2(s))^k$.
That is, we can write 
\be
 \frac{1}{\tau^k}\||\phi^{(k)}(1)\rangle\|\le 
 (2k-1)!!\left(\frac{\|\dot{W}(1)\|}{\tau g^2(1)}\right)^k
 +\cdots+\frac{\|\partial^{k}W(1)\|}{\tau^kg^{k+1}(1)}. \label{phik}
\ee

In the simplest case where $W(s)$ is linear in $s$, 
only the first term remains.
We obtain the naive result in Eq.~(\ref{g2}) as the adiabatic condition.
This conclusion is changed when $\dot{W}(1)=0$.
For example, when $\partial^k W(1)=0$ for $k\ne 2$,
the expansion parameter is $\|\ddot{W}(1)\|/\tau^2 g^{3}(1)$ and
the adiabatic condition is changed to 
\be
 \tau \ge \mathrm{const.}\times\frac{\sqrt{\|\ddot{W}(1)\|}}{g^{3/2}(1)}.
\ee
Generally, many terms contribute to the bound 
and we would find a complicated behavior
$\tau \sim \mathrm{const.}\times g^{-\alpha}(1)$ 
ranging from $\alpha=1$ to $\alpha=2$.
In a regime where $g(1)$ is small, the contribution with $\alpha=2$
gives the worst case bound.

We note that the result is only dependent on quantities
at $s=1$ and is independent of the history of the time evolution.
This is because the relaxation occurs quickly in the present time evolution.

\subsection{Bounds of expansion coefficients (2)}
\label{sec:bound2}

The first term on the right hand side of Eq.~(\ref{phik})
involves a factor $(2k-1)!!$
which becomes very large for $k\to\infty$ and it is not clear
whether the infinite series expansion makes sense.
In rigorous treatments of adiabatic theorems, we take $M$ finite
and keep the last term in Eq.~(\ref{Mth}).

$U_\tau(s,s')$ in the last term
acts on a projected space as $Q(s)U_\tau(s,s')Q(s')$
and involves an exponentially-small factor.
We decompose the integral at $s=1$ as 
\be
 \int_0^1 ds\,U_\tau(1,s)|\dot{\phi}^{(M)}(s)\rangle
 &=& \int_0^{1-\delta}ds\,U_\tau(1,s)|\dot{\phi}^{(M)}(s)\rangle
 \no\\ &&
 +\int_{1-\delta}^1 ds\,U_\tau(1,s)|\dot{\phi}^{(M)}(s)\rangle.
\ee
We set $\frac{1}{\tau g(1)}\ll \delta\ll 1$.
The first term is exponentially suppressed as 
\be
 \int_0^{1-\delta}ds\,e^{-\tau \int_s^1 ds'\, g(s')}
 < \int_0^{1-\delta}ds\,e^{-\tau \int_{1-\delta}^1 ds'\, g(s')}
 \sim (1-\delta)\,e^{-\tau g(1)\delta}\ll 1.
\ee
The second term is evaluated as 
\be
 \int_{1-\delta}^1 ds\,U_\tau(1,s)|\dot{\phi}^{(M)}(s)\rangle
 \sim \int_{1-\delta}^1 ds\,U_\tau(1,1)|\dot{\phi}^{(M)}(1)\rangle
 \sim \delta|\dot{\phi}^{(M)}(1)\rangle.
\ee

Now we obtain
\be
 d_\tau(1) &\sim& \frac{1}{2}\sum_{n=0}^{N-1}
 \left|\sum_{k=1}^M\left(\frac{-1}{\tau}\right)^k \langle n|\phi^{(k)}(1)\rangle
 -\delta\left(\frac{-1}{\tau}\right)^M
 \langle n|\dot{\phi}^{(M)}(1)\rangle\right| 
 \no\\
 &\le& \frac{\sqrt{N}}{2}
 \left(\sum_{k=1}^M\frac{1}{\tau^k}\||\phi^{(k)}(1)\rangle\|
 +\frac{\delta}{\tau^M}
 \||\dot{\phi}^{(M)}(1)\rangle\|\right).
\ee
We examine the simplest case where $W(s)$ is linear in $s$.
In that case, we obtain 
\be
 && \sum_{k=1}^M\frac{1}{\tau^k}\||\phi^{(k)}(1)\rangle\|
 +\frac{\delta}{\tau^M}\||\dot{\phi}^{(M)}(1)\rangle\|
 \no\\
 &=&\sum_{k=1}^M(2k-1)!!\left(\frac{\|\dot{W}(1)\|}{\tau g^2(1)}\right)^k
 +\delta(2M+1)!!\left(\frac{\|\dot{W}(1)\|}{\tau g^2(1)}\right)^{M+1}\tau g(1).
 \label{expm}
\ee

We discuss the condition that the last term
on the right hand side of Eq.~(\ref{expm}) is small.
We optimize $M$ so that this term is minimized~\cite{Elgart12}.
When $M$ is large enough, we can write 
\be
 (2M+1)!!\delta \left(\frac{\|\dot{W}(1)\|}{\tau g^2(1)}\right)^{M+1}\tau g(1)
 \sim \delta\left(\frac{2M\|\dot{W}(1)\|}{\tau g^2(1)}\right)^{M}
 \frac{\|\dot{W}(1)\|}{g(1)},
\ee
and the optimized value $M=M_{\mathrm{opt}}$ is obtained as 
\be
 M_{\mathrm{opt}}\sim\frac{e^{-1}\tau g^2(1)}{2\|\dot{W}(1)\|}.
 \label{mopt}
\ee
Then, we have
\be
 \delta\left(\frac{2M_{\mathrm{opt}}\|\dot{W}(1)\|}{\tau g^2(1)}\right)^{M_{\mathrm{opt}}}
 \frac{\|\dot{W}(1)\|}{g(1)}
 \sim \delta\exp\left(-\frac{e^{-1}\tau g^2(1)}{2\|\dot{W}(1)\|}\right)
 \frac{\|\dot{W}(1)\|}{g(1)}.
\ee
This expression implies that the condition in Eq.~(\ref{g2}) expected
from the naive adiabatic theorem is not justified.
To make this quantity small for $g(1)\to 0$ we need 
\be
 \tau \ge 2e\alpha\frac{\|\dot{W}(1)\|}{g^2(1)}
 \left|\ln\frac{g(1)}{g_0}\right|,  \label{g2log}
\ee
with $\alpha\ge 1$.
$g_0$ represents a proper scale that makes $g(1)/g_0$ dimensionless.

Finally, we show that the first term in Eq.~(\ref{expm})
is negligible when we use Eqs.~(\ref{mopt}) and (\ref{g2log}).
It can be shown as 
\be
&&\sum_{k=1}^{M_{\mathrm{opt}}} (2k-1)!!
\left(\frac{\|\dot{W}(1)\|}{\tau g^2(1)}\right)^k
\no\\
&\sim& \sum_{k=1}^{M_{\mathrm{opt}}}
\left(2k\frac{\|\dot{W}(1)\|}{\tau g^2(1)}\right)^k
\no\\
&\le&
\sum_{k=1}^{\sqrt{M_{\mathrm{opt}}}}
\left(\sqrt{2e^{-1}\frac{\|\dot{W}(1)\|}{\tau g^2(1)}}\right)^k
+\sum_{k=\sqrt{M_{\mathrm{opt}}}+1}^{M_{\mathrm{opt}}}
\left(e^{-1}\right)^k
\no\\
&\le& 2\sqrt{2e^{-1}\frac{\|\dot{W}(1)\|}{\tau g^2(1)}}
+2e^{-\sqrt{M_{\mathrm{opt}}}}.
\ee
In the last line, we use $\sum_{k=1}^M r^k\le 2r$ for $r\ll 1$.
The last expression approaches zero for $g(1)\to 0$.

In conclusion, the distance in Eq.~(\ref{tracedist})
is kept small at $g(1)\to 0$ if, but not only if,
the condition in Eq.~(\ref{g2log}) is satisfied. 
We can apply a similar analysis when $W(s)$ is nonlinear in $s$.
We note that the analysis in the present subsection gives the worst case bound.
The logarithmic factor in Eq.~(\ref{g2log}) becomes large only
when $g(1)$ takes a considerably small value.

\section{Examples}
\label{sec:ex}

\subsection{Two-state system}

\begin{figure}[t]
\begin{center}
\includegraphics[width=0.9\columnwidth]{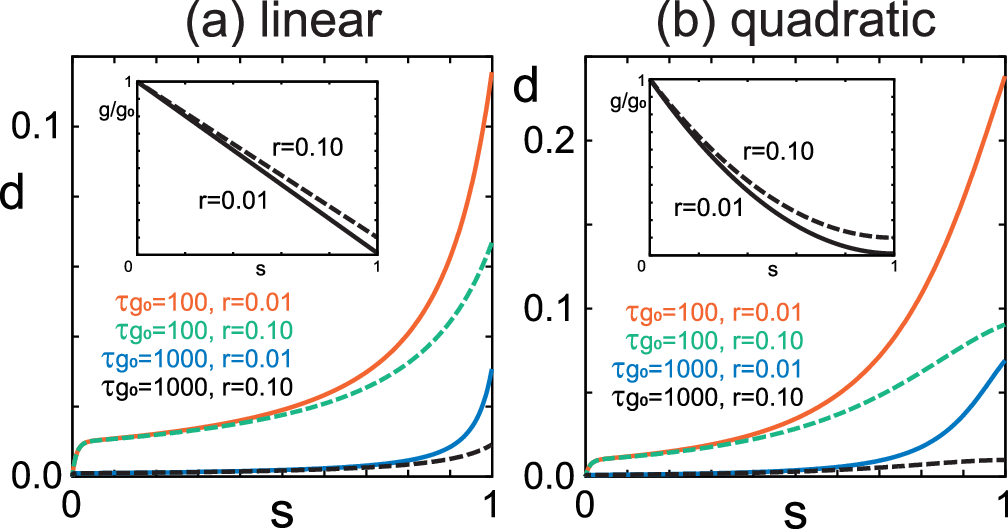}
\caption{
The trace distance $d_\tau(s)$ for a two-state process.
We consider the linear case in the panel (a) and
the quadratic case in (b).
The inset in each panel represents $g(s)$.
}
\label{fig01}
\end{center}
\end{figure}
\begin{figure}[t]
\begin{center}
\includegraphics[width=0.9\columnwidth]{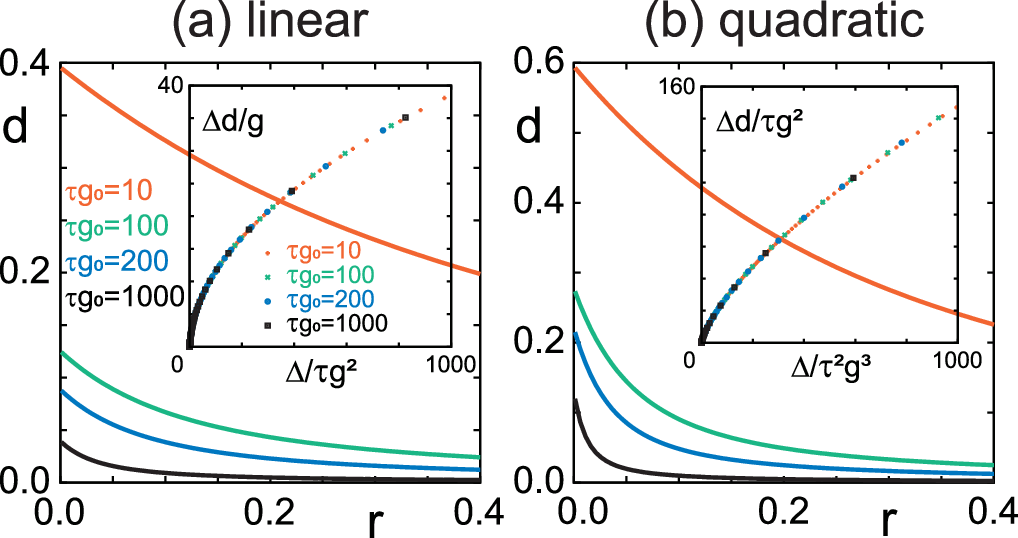}
\caption{
$d_\tau(1)$ as a function of $r=g(1)/g(0)$ 
for the linear case (a) and the quadratic case (b).
The inset in each panel shows that
all curves collapse into a single curve.
We use notations $g=g(1)$, 
$\Delta=|\dot{g}(1)|=g_0(1-r)$ in the panel (a), and 
$\Delta=\ddot{g}(1)=2g_0(1-r)$ in the panel (b).
}
\label{fig02}
\end{center}
\end{figure}
\begin{figure}[t]
\begin{center}
\includegraphics[width=0.9\columnwidth]{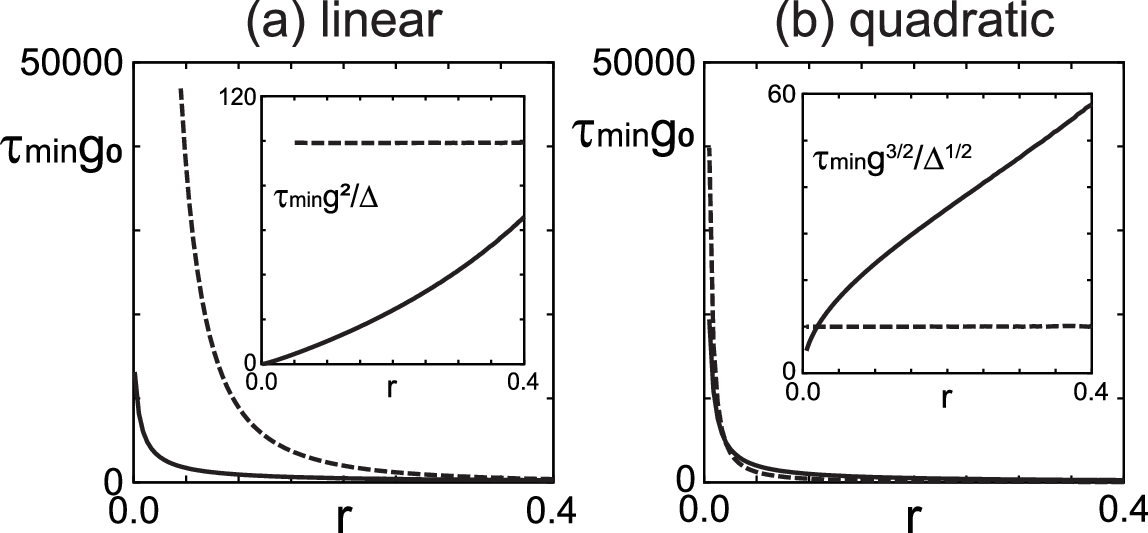}
\caption{
Solid lines represent the minimum annealing time $\tau_{\mathrm{min}}$
that satisfies $d_\tau(1)\le\delta$.
Dashed lines represent $\tau_{\mathrm{min}}$
that satisfies $|\dot{g}(1)|d_\tau(1)/g(1)\le \delta$ for the linear case (a)
and $\ddot{g}(1) d_\tau(1)/\tau g^2(1)\le \delta$ for the quadratic case (b).
We take $\delta=0.01$ and use notations $g=g(1)$, 
$\Delta=|\dot{g}(1)|=g_0(1-r)$ in the panel (a), and 
$\Delta=\ddot{g}(1)=2g_0(1-r)$ in the panel (b).
}
\label{fig03}
\end{center}
\end{figure}

We treat the case at $N=2$.
Although it is a very simple system,
it gives a nontrivial behavior and
we can find various applications in literatures, 
especially in the fields of chemistry and biology.
The adiabatic/nonadiabatic dynamics was discussed
for periodically driven systems giving nontrivial geometrical
results~\cite{Sinitsyn07, Takahashi20}.
In the adiabtaic theorem, we basically intend to change the parameter 
linearly from an initial setting to a different one.

In the two-state processes,
the transition-rate matrix is generally parametrized as 
\be
 W(s)=g(s)\bmat{cc} -(1-p(s)) & p(s) \\ 1-p(s) & -p(s) \emat.
\ee
$p(s)$ represents a probability and the stationary distribution is given by
\be
 |p^{(\mathrm{st})}(s)\rangle = \bmat{c} p(s) \\ 1-p(s) \emat.
\ee
$g(s)$ is positive and represents the decay rate in Eq.~(\ref{gap}).
In the present case, the decay rate appears in the transition-rate matrix
as an overall factor and the stationary state is independent of $g(s)$.

The time-evolved state is written as 
\be
 |p_\tau(s)\rangle-|p^{(\mathrm{st})}(s)\rangle =
 -\int_0^s ds'\,\dot{p}(s')e^{-\tau\int_{s'}^{s} ds''\,g(s'')}\bmat{c} 1 \\ -1 \emat,
\ee
and the asymptotic expansion gives
\be
 d_\tau(1) &=& \left|
 \sum_{k=1}^M\left(-\frac{1}{\tau}\right)^k\left[
 \left.\left(\frac{1}{g(s)}\partial_s\right)^k p(s)\right|_{s=1}
 \right.\right. \no\\ && \left.\left.
 -e^{-\tau \int_0^1 ds\,g(s)}
 \left.\left(\frac{1}{g(s)}\partial_s\right)^k p(s)\right|_{s=0}
 \right]
 \right. \no\\ & & \left.
 -\left(-\frac{1}{\tau}\right)^M\int_0^1 ds\,e^{-\tau \int_s^1 ds\,g(s)}
 \partial_s\left(\frac{1}{g(s)}\partial_s\right)^M p(s)\right|.
 \label{two1}
\ee

In the following, we treat the case $p(s)=s$.
The stationary state changes linearly from 
$|p^{(\mathrm{st})}(0)\rangle = (0,1)^\mathrm{T}$
to $|p^{(\mathrm{st})}(1)\rangle = (1,0)^\mathrm{T}$.
As for the decay rate $g(s)$, we consider the linear and quadratic cases
\be
 g(s)=\left\{\begin{array}{ll}
 \displaystyle
 g_0[1-(1-r)s] & \mbox {linear} \\
 \displaystyle
 g_0[r+(1-r)(1-s)^2] & \mbox{quadratic} \end{array}\right.,
\ee
where $g_0$ and $r$ are positive.
We mainly discuss the domain with $r\ll 1$.
Then, the decay rate decreases as a function of $s$
from $g(0)=g_0$ to $g(1)=g_0r$.
In addition to the standard linear protocol,
we treat the quadratic one.
The latter corresponds to changing the protocol slowly around
the point $s=1$ where the gap becomes minimum.
As we see below, it enhances the performance of the adiabatic operation.
The parameter $g_0$ determines the overall scale of the dynamics
and the result is dependent on $\tau$ and $r$.
The relaxation condition gives $\tau g(1)=\tau g_0r\gg 1$.
We show $d_\tau(s)$ for several values of $\tau$ and $r$
in Fig.~\ref{fig01}.

In the linear case, 
neglecting exponentially-decaying contributions,
we can write Eq.~(\ref{two1}) as
\be
 d_\tau(1) &\sim& \left|
 \sum_{k=1}^M(2k-3)!!\left(-\frac{|\dot{g}(1)|}{\tau g^2(1)}\right)^k
 \frac{g(1)}{|\dot{g}(1)|}
 \right. \no\\ && \left. 
 -(2M-1)!!\int_0^1 ds\,e^{-\tau \int_s^1 ds\,g(s)}
 \left(-\frac{|\dot{g}(1)|}{\tau g^2(s)}\right)^M\right|. \label{two12}
\ee
This expression is slightly different from Eq.~(\ref{expm}).
In the present case, the stationary state is independent of $g(s)$,
which gives a difference from the general argument.
The form of the first term implies that $|\dot{g}(1)|d_\tau(1)/g(1)$
is a function of $|\dot{g}(1)|/\tau g^2(1)$.
In a similar way, we can find the asymptotic form of the quadratic case as 
\be
 d_\tau(1)\sim\left|\frac{\tau g^2(1)}{\ddot{g}(1)}
 \left[\frac{\ddot{g}(1)}{\tau^2g^3(1)}
 -\left(\frac{\ddot{g}(1)}{\tau^2g^3(1)}\right)^2
 +10\left(\frac{\ddot{g}(1)}{\tau^2g^3(1)}\right)^3+\cdots\right]
 \right|. \label{two2}
\ee
This implies that $\ddot{g}(1) d_\tau(1)/\tau g^2(1)$
is a function of $\ddot{g}(1)/\tau^2 g^3(1)$.
We can confirm these expectations in Fig.~\ref{fig02}.

Applying the method discussed in Sec.~\ref{sec:bound2}, 
we find the adiabatic condition
\be
 \tau \ge \left\{\begin{array}{ll}
 \displaystyle
 \mathrm{const.}\times\frac{|\dot{g}(1)|}{g^{2}(1)} & \mbox{linear} \\
 \displaystyle
 \mathrm{const.}\times\frac{|\ddot{g}(1)|^{1/2}}{g^{3/2}(1)} & \mbox{quadratic}
 \end{array}\right..
 \label{scalingtwo}
\ee
We note that a logarithmic factor is not required in this case. 
To discuss the scaling relations, 
we calculate the minimum annealing time $\tau_{\mathrm{min}}$
satisfying the condition $d_\tau(1)\le \delta$ with a small constant $\delta$.
The result is plotted by solid lines in Fig.~\ref{fig03}.
The inset of each panel implies that
the scaling relation is different from Eq.~(\ref{scalingtwo}).
Scaled annealing time $\tau_{\mathrm{min}}g^{\alpha}(1)$
with $\alpha=2$ or $3/2$ approaches zero at $g(1)\to 0$.
This behavior is due to an overall factor 
in Eqs.~(\ref{two12}) and (\ref{two2}).
To remove the effect of the factor, we consider modified conditions
$|\dot{g}(1)|d_\tau(1)/g(1)\le \delta$ for the linear case and 
$\ddot{g}(1) d_\tau(1)/\tau g^2(1)\le \delta$ for the quadratic case.
The result is plotted by dashed lines in Fig.~\ref{fig03}.
The plot in each panel clearly indicates 
$\tau_{\mathrm{min}}g^{\alpha}(1)\sim\mathrm{const.}$ with $\alpha=2$ or $3/2$.
Since the modified condition guarantees $d_\tau(1)\ll 1$, 
we can conclude the scaling relations (\ref{scalingtwo}).

\subsection{Three-state system}

\begin{figure}[t]
\begin{center}
\includegraphics[width=0.5\columnwidth]{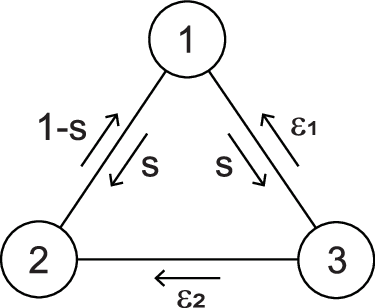}
\caption{
A three-state inhomogeneous process.
The state is basically transferred from the node 1 to 2.
Each transition rate is characterized by the quantity
attached to each arrow.
}
\label{fig04}
\end{center}
\end{figure}

\begin{figure}[t]
\begin{center}
\includegraphics[width=0.5\columnwidth]{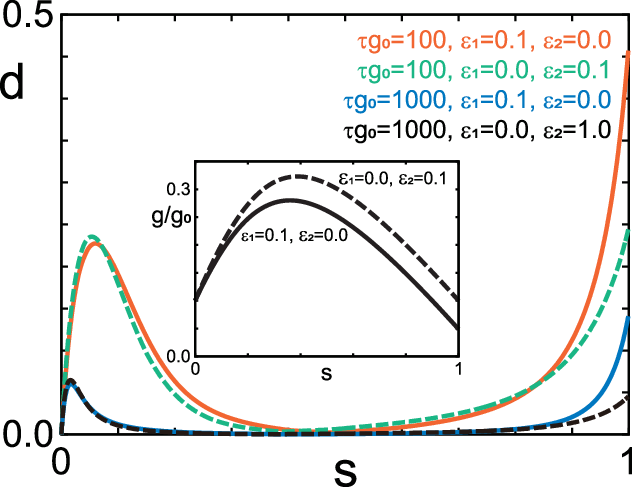}
\caption{
The trace distance $d_\tau(s)$ for a three-state process.
The inset represents $g(s)$.
}
\label{fig05}
\end{center}
\end{figure}

\begin{figure}[t]
\begin{center}
\includegraphics[width=0.9\columnwidth]{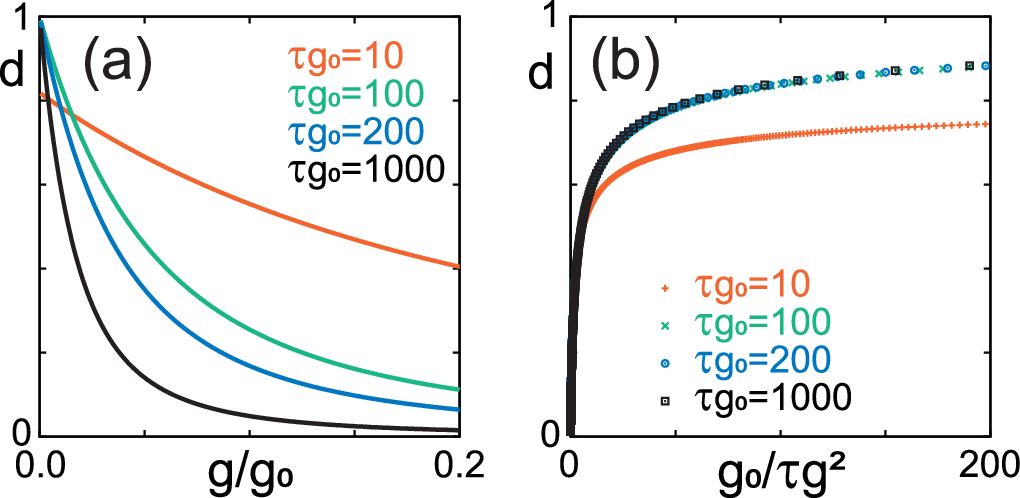}
\caption{
(a) $d_\tau(1)$ as a function of $g=g(1)$.
(b) The same data is plotted as a function of $1/\tau g^2$.
}
\label{fig06}
\end{center}
\end{figure}
\begin{figure}[t]
\begin{center}
\includegraphics[width=0.9\columnwidth]{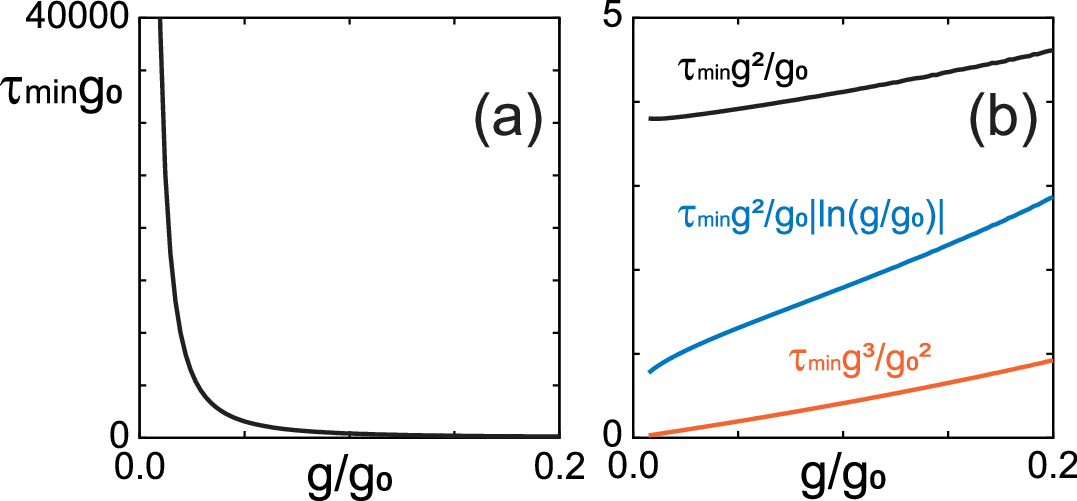}
\caption{
The minimum annealing time $\tau_{\mathrm{min}}$
that satisfies $d_\tau(1)\le\delta$ with $\delta=0.01$.
All curves in the panel (a) and (b) represent the same data.
}
\label{fig07}
\end{center}
\end{figure}

Next, we discuss the case at $N=3$.
This is the second simplest system and
is known as the minimum model that breaks the local detailed balance condition.
An example of the adiabatic operation for the three-state system
can be found in \cite{Astumian07}.

We treat a system depicted in Fig.~\ref{fig04}.
The transition-rate matrix is linear in $s$ and is given by 
\be
 W(s)=g_0\bmat{ccc}
 -2s & 1-s & \epsilon_1 \\
 s & -(1-s) &  \epsilon_2\\
 s & 0 & -(\epsilon_1+\epsilon_2) \emat,
\ee
where $g_0$, $\epsilon_1$, and $\epsilon_2$ are positive. 
The state is in the node 1 at $s=0$ and
is basically transferred to the node 2.
The stationary state is given by 
\be
 |p^{(\mathrm{st})}(s)\rangle=
 \frac{1}{\epsilon_1+\epsilon_2(1+s)+s(1-s)}
 \bmat{c} (\epsilon_1+\epsilon_2)(1-s) \\ (\epsilon_1+2\epsilon_2) s \\
 s(1-s) \emat.
\ee
Transitions to the node 3 give deviations from the stationary state.
The minimum decay rate is given by 
\be
 g(s) &=& g_0\left[\frac{1+s+\epsilon_1+\epsilon_2}{2}
 \right. \no\\ &&\left.
 -\sqrt{
 \left(\frac{1+s+\epsilon_1+\epsilon_2}{2}\right)^2
 -[\epsilon_1+\epsilon_2(1+s)+s(1-s)]}\right].
\ee
This function becomes minimum at $s=1$
as shown in the inset of Fig.~\ref{fig05}.
$g(1)$ takes a small value when $\epsilon_1\ll 1$ and $\epsilon_2\ll 1$.
As we show in Fig.~\ref{fig05}, 
$d_\tau(s)$ deviates from zero when $g(s)$ takes a small value.

Since the effects of $\epsilon_1$ and $\epsilon_2$ give basically similar
results, we set $\epsilon_2=0$ in the following analysis.
Figure~\ref{fig06} shows $d_\tau(1)$ as a function of $g(1)$.
In the present case, the transition-rate matrix is linear in $s$
and we can use Eq.~(\ref{expm}).
When $d_\tau(1)$ is plotted as a function of $1/\tau g^2(1)$,
all curves collapse into a single curve if $\tau g(1)$ is large enough
as we see in the panel (b) of Fig.~\ref{fig06}.

Figure~\ref{fig07} shows the minimum annealing time $\tau_{\mathrm{min}}$
that satisfies $d_\tau(1)\le \delta$.
The plots in the panel (b) show that the scaling
$\tau_{\mathrm{min}}\sim 1/g^2(1)$ holds approximately.
Since the general discussion uses several inequality relations, 
it is generally difficult to fit the data by the scaling of the upper bound.

\section{Conclusion}
\label{sec:conc}

In conclusion, we have developed the adiabatic theorem for
classical stochastic processes.
The formulation goes along the theorem for quantum systems
and we find a great simplification due to several properties of
the probability distribution.
Since the classical master equation describes relaxation
to the stationary state,
the time-evolved state is insensitive to the whole history
of the time evolution.
As a result, the theorem can be described basically
by using quantities defined instantaneously.

By using a rigorous treatment, we find Eq.~(\ref{g2log}) 
as a worst case bound.
It is a sufficient condition and 
the logarithmic factor become large only when the minimum decay rate takes
a considerably small value.
In fact, we found in several examples
that the naive scaling in Eq.~(\ref{g2}) is enough
to guarantee the error suppression.

In the present study, we treated the generic form of
the classical master equation.
One of most interesting applications is simulated annealing where
the stationary distribution is given by the Gibbs distribution
of an Ising Hamiltonian.
To reproduce the Geman--Geman formula~\cite{Geman84}  
from the quantum-classical mapping~\cite{Somma07},
we need to discuss the energy gap of the effective Hamiltonian.
Combining with the condition in Eq.~(\ref{g2}), 
we can find a time dependence of the temperature~\cite{Morita08, Kimura22}.
It is not clear how the result is changed when
the condition in Eq.~(\ref{g2}) is modified
and we leave the problem as an open one.

\section*{Acknowledgments}
The author is grateful to Yasuhiro Utsumi for useful discussions.
This work was supported by 
JSPS KAKENHI Grants No. JP20H05666 and No. JP20H01827.


\end{document}